\documentclass[12pt]{article}

\usepackage{graphicx}

\title{Regression on a Graph}
\author{Arne Kovac and Andrew D.A.C. Smith}
\date{}

\newtheorem{theorem}{Theorem}

\begin{document}

\begin{titlepage}

\thispagestyle{empty}

\maketitle

\begin{abstract}
The `Signal plus Noise' model for nonparametric regression can be extended to the case of observations taken at the vertices of a graph.
This model includes many familiar regression problems.
This article discusses the use of the edges of a graph to measure roughness in penalized regression.
Distance between estimate and observation is measured at every vertex in the $L_2$ norm, and roughness is penalized on every edge in the $L_1$ norm.
Thus the ideas of total-variation penalization can be extended to a graph.
The resulting minimization problem presents special computational challenges, so we describe a new, fast algorithm and demonstrate its use with examples.

Further examples include a graphical approach that gives an improved estimate of the baseline in spectroscopic analysis, and a simulation applicable to discrete spatial variation.
In our example, penalized regression outperforms kernel smoothing in terms of identifying local extreme values.
In all examples we use fully automatic procedures for setting the smoothing parameters.
\end{abstract}

\end{titlepage}

\section{Introduction}
There are a number of statistical models that contain some sort of graphical structure. Examples include image analysis, disease risk mapping and discrete spatial variation. We focus on those for which penalized regression is appropriate, and can be thought of in terms of the `signal + noise' framework.

We consider the regression of a continuous response variable on one or more explanatory covariates. Often there is some sort of graphical structure in and between the observations, or some obvious neighbouring scheme that gives rise to a graph.
We think of the locations of the observations as the vertices of the graph. The edges may be suggested by the neighbouring scheme or by the covariate values.
We will see some examples in this section.

A model for data on the graph $(\mathcal{V},\mathcal{E})$, which has vertices in the set $\mathcal{V}$ and edges in the set $\mathcal{E}$, is
\begin{displaymath}
\begin{array}{cccccr}
\mbox{Data} & = & \mbox{Signal} & + & \mbox{Noise} \\
y_i & = & f_i & + & \sigma z_i, & \quad i \in \mathcal{V}.
\end{array}
\end{displaymath}
The noise terms, $z_i$, are usually assumed to be independent realizations of a random variable with zero mean and unit variance. Under this model regression on a graph involves estimating the underlying signal values $f_i$, for all vertices $i$ in the set $\mathcal{V}$. 
We use the edges to measure the complexity of the estimate.

\begin{figure}
\begin{center}
\includegraphics{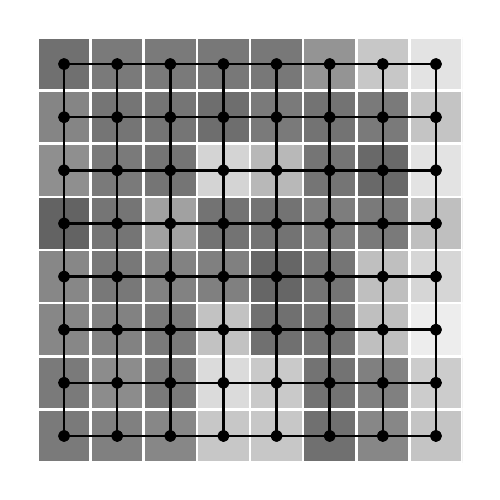}
\includegraphics{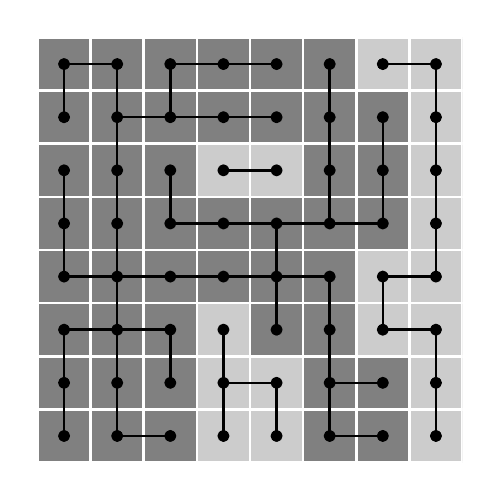}
\caption{\label{graph}Example of a graphical structure present in a regression situation. 
The noisy image (left) shows a suitable graph for regression, based on the 4-neighbourhood. 
On the noiseless version (right) only the edges in the active set are shown.}
\end{center}
\end{figure}

Figure~\ref{graph} shows an example of regression on a graph: a small, noisy image with 64 pixels. The responses are the grey levels of the pixels, so each pixel is a vertex of the graph. A natural choice of edges connects each pixel with its neighbours, resulting in the graph superimposed on the left-hand image in Figure~\ref{graph}. Regression on this graph involves estimating the underlying signal image, which is displayed in the right-hand image.

In this article we discuss penalized regression on the graph $(\mathcal{V},\mathcal{E})$. Penalized regression fits an estimate that is close to the data, but penalises rough or complicated estimates. With an observation at every vertex, we can measure the distance between observed and estimated values by the sum of the distances at each vertex. The complexity of the estimate can be measured by the differences between the estimated values at adjacent observations. This measurement is therefore the sum of absolute differences at each edge.

We discuss the penalized regression estimate that minimises
\begin{displaymath}
Q(f) := \frac{1}{2} \sum_{i \in \mathcal{V}} w_i (f_i - y_i)^2 + \sum_{(i,j) \in \mathcal{E}} \lambda_{i,j} |f_j - f_i|
\end{displaymath}
for given weights $w_i \geq 0$, for $i \in \mathcal{V}$, and smoothing parameters $\lambda_{i,j} > 0$, for $(i,j) \in \mathcal{E}$. This is the sum of a term that penalises distance from the data plus a term that penalises roughness. The first term is the distance from the data, measured at every vertex in the $L_2$ norm. The second term is the weighted sum of roughness at every edge, measured in the $L_1$ norm. Our model allows for a different weight or smoothing parameter at each vertex and each edge.

Although it is usual, in graph theory, to denote the edges by unordered pairs, we will treat $\mathcal{E}$ as a set of ordered pairs for convenience of notation. This does not mean that $(\mathcal{V},\mathcal{E})$ is a directed graph, since the ordering can be completely arbitrary. We do, however, consider there to be at most one edge that joins any pair of vertices. This is because it makes no sense to split the penalty between two vertices over more than one edge. 

\subsection{Motivating examples}
As a first motivating example, we consider the problem of nonparametric regression between two continuous variables. 
Suppose we have response observations $y_1, \ldots, y_n$ taken at strictly ordered design points.
There is a natural neighbouring structure: the first observation is adjacent to the second, the second is next to the third, and so on. 
Hence a natural graphical structure for this example is given by $(\mathcal{V}_2,\mathcal{E}_2)$, where
\begin{displaymath}
\mathcal{V}_2 = \left\lbrace 1,2,\ldots,n \right\rbrace
\mbox{ and }
\mathcal{E}_2 = \left\lbrace (1,2), (2,3), \ldots, (n-1,n) \right\rbrace.
\end{displaymath}

The minimization of $Q(f)$ provides an estimate of $f_i$ at every observation.
If we let $w_i = 1$ for all $i \in \mathcal{V}_2$ and use the convenient shorthand $\lambda_i = \lambda_{i,i+1}$, then $Q(f)$ becomes
\begin{equation}
\label{eq:tvdenoising}
\frac{1}{2} \sum_{i=1}^n (y_i - f_i)^2 + \sum_{i=1}^{n-1} \lambda_i |f_{i+1} - f_i|
\end{equation}
and the roughness penalty is the weighted total variation of the estimate.

Total variation can be extended to higher dimensions to tackle, for example, image analysis. An image can be thought of as an $n_1 \times n_2$ grid of pixels, with observations at each pixel. Then the set of vertices of the graph is the set of pixels
\begin{displaymath}
\mathcal{V}_4 = \left\lbrace (i_1,i_2) : i_1=1,\ldots,n_1 , i_2=1,\ldots,n_2 \right\rbrace.
\end{displaymath}

There are a number of neighbouring structures in use in image analysis. The simplest is the 4-neighbourhood (Winkler 2003, p.\ 57) in which a pixel has neighbours immediately above, below, to the left and to the right. This neighbouring scheme suggests the set of edges
\begin{displaymath}
\mathcal{E}_4 = 
\left\lbrace ( (i_1,i_2),(i_1,i_2+1) ) \in \mathcal{V}_4^2 \right\rbrace \cup
\left\lbrace ( (i_1,i_2),(i_1+1,i_2) ) \in \mathcal{V}_4^2 \right\rbrace.
\end{displaymath}
Figure~\ref{graph} shows a picture of this graph.

Using the graph $(\mathcal{V}_4, \mathcal{E}_4)$, we can find a denoised image by minimising $Q(f)$. 
Now the roughness penalty is a measure of the total variation in the horizontal direction plus the total variation in the vertical direction.

\subsection{Review of existing methods}
Mammen and van de Geer (1997) first discussed the estimator obtained by minimising (\ref{eq:tvdenoising}) where $\lambda$ is a global smoothing parameter.
Some authors have allowed the smoothing parameters to differ. For example Davies and Kovac (2001) alter them during their local squeezing procedure.
There are fast algorithms that find the solution to this specific minimization problem, in particular the taut string algorithm of Davies and Kovac (2001), which has $O(n)$ computational complexity.

The estimator that minimises (\ref{eq:tvdenoising}), in which error is measured in the $L_2$ norm and roughness in the $L_1$ norm, is a nonparametric version of the least absolute shrinkage (Lasso) estimator (Tibshirani 1996). Therefore the estimator that minimises $Q(f)$ can be seen as a generalization of the nonparametric Lasso to any graph.
There are other methods of penalized regression, with different roughness measures, that have been applied to observations on a graph. Belkin et al.\ (2004) describe an algorithm for Tikhonov regularization. Their algorithm measures roughness at every edge in the $L_2$ norm.

Koenker and Mizera (2004) employ a penalty term for triograms. Given irreg\-ularly-spaced observations, they create a graph by computing a Delaunay triangulation of the observations. Their penalty term is also a weighted sum over all edges of the triangulation. However they measure roughness as the squared ($L_2$) differences between gradients.
Jansen et al.\ (2009) have discussed wavelet lifting as a method for regression on a graph. Like Koenker and Mizera, the authors use a Delaunay triangulation.

Our algorithm is based on ideas similar to active set methods, which features in a number of algorithms, including that of Goldfarb and Idnani (1983).

\section{Optimization Algorithm}
In Theorem~\ref{theorem:one} below we give a sufficient condition for $f$ to minimize $Q(f)$ and in Subsection \ref{section:algorithm} we present a fast algorithm for finding such a minimizer. 
The minimum exists because $Q(f)$, as a sum of convex functions, is convex itself. 
Therefore any local minimum of $Q(f)$ will be a global minimum, and the set of all global minima will be a convex set.
In the important case where all the weights $w_i$ are strictly positive a unique global minimum exists, because $Q(f)$ is strictly convex.

\subsection{Sufficient condition for minimization}
The solution to the minimization problem is characterized by \emph{regions of constant value}, that is, sets of neighbouring vertices that share the same value of $f$. We define such regions by use of a special \emph{active set} of edges, indexed by $\mathcal{A}$. This consists of edges $(i,j) \in \mathcal{E}$ for which $f_i = f_j$, such that the graph $(\mathcal{V},\mathcal{A})$ is acyclic.
Note that, unlike the definition of active set used in many optimization algorithms, there can still be edges $(i,j) \notin \mathcal{A}$ such that $f_i = f_j$.

We will denote by $\mathcal{R}(k)$ the entire region of constant value that contains the vertex $k$. 
More formally let
\begin{displaymath}
\mathcal{R}(k) = \left\lbrace i \in \mathcal{V} : \mbox{$i$ is connected to $k$ in $(\mathcal{V},\mathcal{A})$} \right\rbrace.
\end{displaymath}
We will also denote by $\mathcal{A}(k)$ that subset of the active set that holds the region $\mathcal{R}(k)$ together, so
\begin{displaymath}
\mathcal{A}(k) = \left\lbrace (i,j) \in \mathcal{A} : i \in \mathcal{R}(k), j \in \mathcal{R}(k) \right\rbrace.
\end{displaymath}
Figure~\ref{graph} shows an example of an active set in the graph $(\mathcal{V}_4, \mathcal{E}_4)$. Note how the edges in the active set join together vertices that share the same value, thus holding together regions of constant value.

Since $(\mathcal{V},\mathcal{A})$ is acyclic, the graph $(\mathcal{R}(k),\mathcal{A}(k))$ is a connected, acyclic graph. This feature is crucial as it allows the region $\mathcal{R}(k)$ to be split into two subregions by removing just one edge $(I,J)$ from $\mathcal{A}(k)$. We will denote these two subregions by $\mathcal{R}(I,J)$ and $\mathcal{R}(J,I)$, where
\begin{displaymath}
\mathcal{R}(I,J) = \left\lbrace i \in \mathcal{R}(I) : \mbox{$i$ is connected to $I$ in $(\mathcal{V},\mathcal{A} \setminus (I,J))$} \right\rbrace.
\end{displaymath}
We associate with the region or subregion $\mathcal{R}(a)$ (where $a = k$ or $a = I,J$) the quantities
\begin{displaymath}
m_a = \sum_{i \in \mathcal{R}(a)} \left( w_i y_i + \sum_{j : (i,j) \in \mathcal{E}} c_{i,j} \lambda_{i,j} - \sum_{j : (j,i) \in \mathcal{E}} c_{j,i} \lambda_{j,i} \right)
\mbox{ and } 
u_a = \sum_{i \in \mathcal{R}(a)} w_i.
\end{displaymath} 
\\

\begin{theorem}
\label{theorem:one}
Suppose there exists a fit $f$ and set of edges $\mathcal{A}$ such that $f_i = f_j$ for all $(i,j) \in \mathcal{A}$ and $(\mathcal{V}, \mathcal{A})$ is acyclic.
Also suppose there are values $c_{i,j}$ such that 
\begin{eqnarray}
& c_{i,j} = \mathop{\mathrm{sign}}(f_j-f_i) \mbox{ or } f_i = f_j \mbox{ for all } (i,j) \in \mathcal{E}, \label{eq:stoppingrule} & \\
& c_{i,j} = \pm 1 \mbox{ if } (i,j) \in \mathcal{A}, & \label{eq:nonzero} \\
& u_k f_k = m_k \mbox{ for all } k \in \mathcal{V}, & \label{eq:entirecondition} \\
\mbox{and} & 
\left| u_{I,J} f_I - (m_{I,J} - c_{I,J} \lambda_{I,J}) \right| \leq \lambda_{I,J} \mbox{ for all } (I,J) \in \mathcal{A}. & \label{eq:splitcondition}
\end{eqnarray}
Then $f$ minimises $Q(f)$.
\end{theorem}
A proof is given in the Appendix.

These conditions can be shown to be similar to the taut string of Davies and Kovac (2001).
When the graph is $(\mathcal{V}_2,\mathcal{E}_2)$ the condition (\ref{eq:splitcondition}) describes a tube and (\ref{eq:entirecondition}) describes a string threaded through the tube and pulled taut (Mammen and van de Geer 1997).

\subsection{Algorithm}
\label{section:algorithm}
The algorithm that we describe can be considered to search for the graph $(\mathcal{V}, \mathcal{A})$ and vector $c$ described in Theorem~\ref{theorem:one}. At any point during the algorithm the current value of $c$ defines a working objective function
\begin{displaymath}
Q(f;c) := \frac{1}{2} \sum_{i \in \mathcal{V}} w_i (f_i - y_i)^2 + \sum_{(i,j) \in \mathcal{E}} |c_{i,j}| \lambda_{i,j} |f_j - f_i|.
\end{displaymath}
When we have satisfied the constraints (\ref{eq:stoppingrule}) then $Q(f;c) = Q(f)$. The current value of $f$ always minimises $Q(f;c)$ so when (\ref{eq:stoppingrule}) holds it also minimises $Q(f)$. For $f$ to minimize $Q(f;c)$ a slightly modified version of Theorem~\ref{theorem:one} tells us that we must have
\begin{equation}
\label{eq:weakcondition}
\mathop{\mathrm{sign}}(c_{i,j}) = \mathop{\mathrm{sign}}(f_j - f_i) \mbox{ when } f_i \neq f_j,
\end{equation}
(\ref{eq:nonzero}) and (\ref{eq:entirecondition}) must hold, and
\begin{equation}
\label{eq:modifiedsplit}
0 \leq - \mathop{\mathrm{sign}}(c_{I,J})(u_{I,J} f_I - m_{I,J}) \leq 2 |c_{I,J}| \lambda_{I,J} \mbox{ for all } (I,J) \in \mathcal{A}.
\end{equation}

We start with $c = 0$.
In this initial case $Q(f;c) = \frac{1}{2} \sum_{i \in \mathcal{V}} w_i (f_i - y_i)^2$, so we start with $f=y$ as this is the minimizer.
Our algorithm gradually increases the penalty on each edge: at each iteration $c_{k,l}$ moves from $0$ to $\pm 1$ for one particular edge $(k,l) \in \mathcal{E}$.
Once (\ref{eq:stoppingrule}) is satisfied for an edge, then it remains satisfied. 
The algorithm stops when (\ref{eq:stoppingrule}) is satisfied at all edges.
This event will occur in a finite time, as stated by Theorem~\ref{theorem:two} below.

\begin{theorem}
\label{theorem:two}
The algorithm described here will terminate in a finite time, and finds a minimizer of $Q(f)$, for any graph, data, weights and smoothing parameters.
\end{theorem}
The proof is contained in the Appendix. \\

We now give precise details about each iteration of our algorithm.
At each iteration we start with $f$ that minimises $Q(\, \cdot \, ; c)$ and move to $\tilde{f} = f + \Delta f$ that minimises $Q(\, \cdot \, ; c+ \Delta c)$.
We start each iteration with an edge $(k,l)$ chosen such that $f_k \neq f_l$ and $c_{k,l} \neq \mathop{\mathrm{sign}}(f_l-f_k)$. We want to move in the direction that satisfies $c_{k,l} = \mathop{\mathrm{sign}}(\tilde{f}_l-\tilde{f}_k)$. The condition (\ref{eq:entirecondition}) tells us we need  $\Delta c$ such that $\Delta c_{i,j} = 0$ for $(i,j) \neq (k,l)$ and $\Delta c_{k,l} = u_k \Delta f_k / \lambda_{k,l}$. 

It is clear that as $c$ changes, $f$ must change to compensate. 
As $c_{k,l}$ changes, the penalty on the edge $(k,l)$ increases, so we must reduce $|f_l-f_k|$ in order to move to the minimum of $Q(\, \cdot \, ; c + \Delta c)$. 

This change must take place within the constraints of the active set. 
Therefore we must alter $f_k$ and $f_l$ uniformly on the whole of the regions $\mathcal{R}(k)$ and $\mathcal{R}(l)$. 
This means $\tilde{f}_i = f_i + \Delta f_k$ for $i \in \mathcal{R}(k)$ and $\tilde{f}_i = f_i + \Delta f_l$ for $i \in \mathcal{R}(l)$. In order to preserve (\ref{eq:entirecondition}) we must have $\tilde{f}_i = f_i$ for $i \in \mathcal{R}(k) \cup \mathcal{R}(l)$.
So the regions $\mathcal{R}(k)$ and $\mathcal{R}(l)$ will move closer together in value.

As the regions move closer together there may need to be changes to the active set. To make sure that these changes happen we will increase the penalty on $(k,l)$ in small steps. 
Specifically we will change $f$ and $c$ only by enough to trigger the first change in the active set.

In this subsection we will discuss the possible changes to the active set as $\mathcal{R}(k)$ and $\mathcal{R}(l)$ move closer together. There are four possible events that could happen: no change, merging of $\mathcal{R}(k)$ and $\mathcal{R}(l)$, amalgamation with a neighbouring region, and splitting a region. 

For each of these events we give, below, the associated values of $\Delta f_k$, $\Delta f_l$ and $\Delta c_{k,l}$. We also describe appropriate adjustments to the active set. In order to trigger the first change in the active set, the algorithm chooses the event for which $|\Delta f_k|$ and $|\Delta f_l|$ are both smallest.
The Appendix contains proofs of these values.

Once the no change or merging steps are complete, we can set $c_{k,l} = \mathop{\mathrm{sign}}(f_l - f_k)$ and the iteration is over.
We choose another edge $(k,l)$ for which $f_k \neq f_l$ and $c_{k,l} \neq \mathop{\mathrm{sign}}(f_l - f_k)$ and iterate again. If there is no such edge then the algorithm stops, since $Q(f;c) = Q(f)$.
Once amalgamation or splitting has taken place, we proceed to further reduce $|f_l - f_k|$, now altering $f$ uniformly on a changed region.

\subsubsection{No change to active set}
There may be no disruption necessary to the active set before $c_{k,l} + \Delta c_{k,l} = \mathop{\mathrm{sign}}(\tilde{f}_l - \tilde{f}_k)$ is satisfied. This means that we have $u_k \tilde{f}_k = m_k$ and $u_l \tilde{f}_l = m_l$, and (\ref{eq:modifiedsplit}) still holds for all $(I,J) \in \mathcal{A}(k) \cup \mathcal{A}(l)$.

This event can only occur if $u_k > 0$ and $u_l > 0$. The associated changes in $f_k$ and $f_l$ are
\begin{displaymath}
\Delta f_k = \frac{(\mathop{\mathrm{sign}}(f_l - f_k) - c_{k,l}) \lambda_{k,l}}{u_k}
\mbox{ and }
\Delta f_l = \frac{(\mathop{\mathrm{sign}}(f_k - f_l) + c_{k,l}) \lambda_{k,l}}{u_l}.
\end{displaymath}

\subsubsection{Merging of the two regions}
Before we reach the target value of $c_{k,l} = \mathop{\mathrm{sign}}(f_l-f_k)$, the regions $\mathcal{R}(k)$ and $\mathcal{R}(l)$ might meet each other in value. This would mean that $\tilde{f}_k=\tilde{f}_l$ and $|f_l-f_k|$ can be decreased no further. 
The changes in $f_k$ and $f_l$ are
\begin{equation}
\label{eq:merge}
\Delta f_k = \frac{u_l}{u_k + u_l} (f_l - f_k)
\mbox{ and }
\Delta f_l = \frac{u_k}{u_k + u_l} (f_k - f_l).
\end{equation}
If $u_k = u_l = 0$ then we can choose $\Delta f_k = (f_l - f_k) / 2$ and $\Delta f_l = (f_k - f_l) / 2$.

Since we now have $f_k = f_l$ we merge the two regions $\mathcal{R}(k)$ and $\mathcal{R}(l)$ by adding $(k,l)$ to the active set.
If there are other edges that join $\mathcal{R}(k)$ and $\mathcal{R}(l)$, then they will not be added to $\mathcal{A}$, even though they share the same value of $f$. This will ensure that the graph $(\mathcal{V},\mathcal{A})$ remains acyclic.

\subsubsection{Amalgamation of a neighbouring region}
Before we reach the minimizer of $Q(\, \cdot \, ; c + \Delta c)$, the value of $f$ in the region $\mathcal{R}(k)$ may meet the value in a neighbouring region that is not $\mathcal{R}(l)$. More formally there may be a vertex $i \in \mathcal{R}(k)$ and $K \notin \mathcal{R}(k) \cup \mathcal{R}(l)$ for which $c_{i,K} \neq 0$ or $c_{K,i} \neq 0$, and $f_k \leq f_K < f_l$ or $f_k \geq f_K > f_l$.

This event is only possible if $u_l > 0$, or if $u_l = u_k = 0$, or if $u_l = 0$ and $f_K = f_k$.
The changes to $f$ associated with this event are
\begin{equation}
\label{eq:amalg}
\Delta f_k = f_K - f_k
\mbox{ and }
\Delta f_l = \left\lbrace \begin{array}{cl} u_k (f_k - f_K) / u_l & u_l > 0, \\
0 & \mbox{otherwise.} \end{array} \right.
\end{equation}

We now have $f_i = f_K$ and if we proceed to alter $f$ we may break the constraint (\ref{eq:weakcondition}) at the edge $(i,K)$ or $(K,i)$. Therefore, if $\mathop{\mathrm{sign}}(c_{i,K}) = \mathop{\mathrm{sign}}(\Delta f_k)$ or $\mathop{\mathrm{sign}}(c_{K,i}) = -\mathop{\mathrm{sign}}(\Delta f_k)$, we add this edge to the active set. 
This will amalgamate the region $\mathcal{R}(K)$ into $\mathcal{R}(k)$.

If there are other edges that join $\mathcal{R}(k)$ and $\mathcal{R}(K)$ then they will not be added to $\mathcal{A}$. This ensures that the graph $(\mathcal{V},\mathcal{A})$ remains acyclic. 
Of course a similar amalgamation might occur with a neighbour of $\mathcal{R}(l)$. 

\subsubsection{Splitting a region}
Before arriving at the minimizer of $Q(\, \cdot \, ; c + \Delta c)$ we must test whether an edge $(I,J) \in \mathcal{A}(k) \cup \mathcal{A}(l)$ should be removed from the active set.
This will split the region $\mathcal{R}(k)$ or $\mathcal{R}(l)$ into two subregions.
If the split takes place it may be necessary to swap the sign of $c_{I,J}$, in order to preserve the constraint (\ref{eq:weakcondition}) at $(I,J)$. 
This will not affect $Q(f;c)$.
We use condition (\ref{eq:modifiedsplit}) to tell us when an edge should be removed, once we have accounted for the possible sign change.

This event can only occur if $u_k > 0$ and $u_l > 0$. The values of $f$ and $c$ at which $(I,J) \in \mathcal{A}(k)$ should be removed are given by
\begin{displaymath}
\Delta f_k = \frac{m_{I,J} - u_{I,J} f_k - c_{I,J} \lambda_{I,J} \pm \mathop{\mathrm{sign}}(f_l - f_k) |c_{I,J}| \lambda_{I,J}}{u_{I,J}}
\mbox{ and }
\Delta f_l = - \frac{u_k}{u_l} \Delta f_k,
\end{displaymath}
with $+$ for $k \in \mathcal{R}(J,I)$ and $-$ for $k \in \mathcal{R}(I,J)$.
The corresponding values for $(I,J) \in \mathcal{A}(l)$ are obtained by swapping $k$ and $l$. 

\section{Computational Complexity}
We now discuss the computational complexity of our algorithm in the setting of
image analysis, in which the graph is $(\mathcal{V}_4, \mathcal{E}_4)$. 
For the sake of simplicity we consider a square image, letting $\mathcal{V}_4$ be an $\eta \times \eta$ grid of vertices. We are interested in the computational complexity in terms of the number of observations, or vertices, $n$. So $n = \eta^2$ and the set $\mathcal{E}_4$ contains $2n - 2n^{1/2}$ edges.

Suppose we were to use a generic active set method to minimize $Q(f;c)$ subject to (\ref{eq:stoppingrule}). This would be very computationally expensive, mainly because we may need to try all possible combinations of $c$ in $\{-1,1\}^{2n - 2n^{1/2}}$, which leads to exponential complexity.
Our algorithm does not need to try all combinations of $c$. In fact once $c_{k,l} = \mathop{\mathrm{sign}}(f_l - f_k)$ is satisfied it will remain satisfied until our algorithm stops. Therefore we only have to consider each edge once when satisfying (\ref{eq:stoppingrule}). So we need only perform $O(n)$ iterations instead of $O(2^{2n - 2n^{1/2}})$.

In addition, our algorithm does not need to check all possible active sets every time we add an edge. In the process of satisfying (\ref{eq:stoppingrule}) for one edge we may need to change the active set many times, through repeated splitting or amalgamation. Since $|f_l - f_k|$ decreases monotonically, once an edge has been removed from $\mathcal{A}(k)$ or $\mathcal{A}(l)$ it cannot be included again during this iteration.
Therefore, during one iteration, every edge may be added once, and removed once, from the active set.
So our algorithm considers at most $2(2n -2n^{1/2})+1$ active sets per iteration.

Finally, for each of these active sets we will need to make some calculations. It is possible to calculate $u_k$, $m_k$ and $u_{I,J}$, $m_{I,J}$ for all $(I,J) \in \mathcal{A}(k) \cup \mathcal{A}(l)$ without visiting a vertex in $\mathcal{R}(k) \cup \mathcal{R}(l)$ more than twice. The algorithm must check for possible neighbouring regions to amalgamate with. It must also check condition (\ref{eq:modifiedsplit}) at every edge in $\mathcal{A}(k)$ and $\mathcal{A}(l)$. Since $(\mathcal{R}(k),\mathcal{A}(k))$ and $(\mathcal{R}(l),\mathcal{A}(l))$ are connected, acyclic graphs, there will only be $|\mathcal{R}(k)| - 1$ and $|\mathcal{R}(l)|-1$ edges to check. Therefore the complexity of the calculation is $O(|\mathcal{R}(k)|+|\mathcal{R}(l)|)$. This is at most $O(n)$, compared with $O(n^3)$ for methods based on matrix inversion, such as that of Goldfarb and Idnani (1983).

We can reduce the computational complexity even further by working with small sub-images that gradually increase in size. We control the order in which the edge constraints (\ref{eq:stoppingrule}) are satisfied in order to keep $|\mathcal{R}(k)|$ and $|\mathcal{R}(l)|$ as small as possible. Here we describe an implementation of our algorithm in which the maximum size of a region grows dyadically.
For the sake of simplicity we will consider $\eta$ to be an integer power of 2. It is easy to adapt this method for other values of $\eta$, and for non-square images.

The edge constraints are satisfied in stages, there being $\log_2 \eta$ stages in total. At stage $p$ we consider those edges in the set
\begin{displaymath}
\left\lbrace ( (i, 2^p q - 2^{p-1}),(i, 2^p q - 2^{p-1} + 1) ) \in \mathcal{E}_4 : q = 1,\ldots,\eta/2^p \right\rbrace
\end{displaymath}
followed by those in the set
\begin{displaymath}
\left\lbrace ( (2^p q - 2^{p-1}, i), (2^p q - 2^{p-1} + 1, i) ) \in \mathcal{E}_4 : q=1,\ldots,\eta/2^p \right\rbrace.
\end{displaymath}

The effect is that as the edges are considered the graph of satisfied edges grows dyadically. At the first stage the graph looks like pairs of vertices, followed by squares of $2 \times 2$ vertices. At the second stage the graph looks like connected rectangles of $2 \times 4$ vertices, followed by squares of $4 \times 4$ vertices. The process continues until all edges are satisfied and the whole square of $\eta \times \eta$ vertices are connected.

The advantage of this implementation is our algorithm will never allow an edge $(k,l)$ in the active set if $c_{k,l} = 0$. Therefore $\mathcal{R}(k)$ and $\mathcal{R}(l)$ can never be larger than the rectangle connected by satisfied edges that contains $k$ and $l$. At stage $p$ this rectangle will contain at most $2^{2p}$ vertices. Furthermore in the process of satisfying $c_{k,l} = \mathop{\mathrm{sign}}(f_l - f_k)$, the active set will only change on edges inside this connected rectangle. So there are at most $2(2^{2p+1} - 2^{p+1})+1$ active sets to consider.

It is possible to find the total computational complexity of this implementation. At every stage we must satisfy constraints on $O(\eta^2 2^{-p})$ edges. For each of these edges we may have to check $O(2^{2p})$ active sets and for each active set perform $O(2^{2p})$ calculations. Therefore the overall complexity is 
\begin{displaymath}
O \left( \sum_{p=1}^{\log_2 \eta} \eta^2 2^{-p} 2^{2p} 2^{2p} \right) = O(\eta^5) = O(n^{5/2}).
\end{displaymath}

\section{Examples}

\subsection{Achieving a constant baseline}
\begin{figure}
\begin{center}
\includegraphics{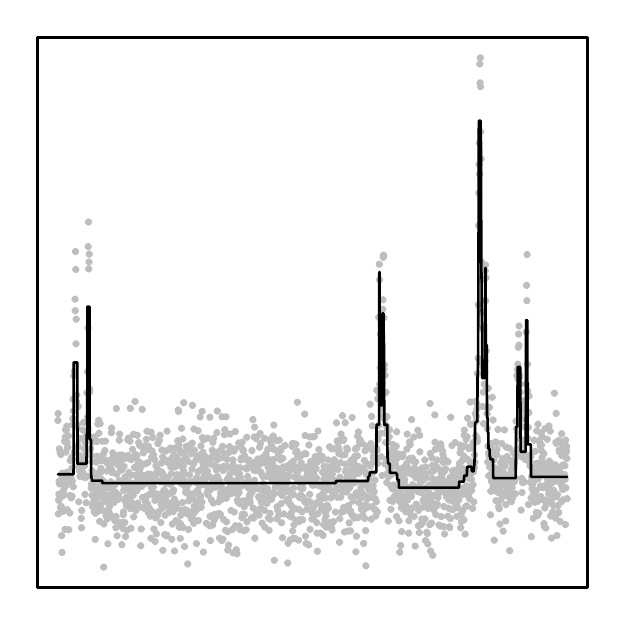}
\includegraphics{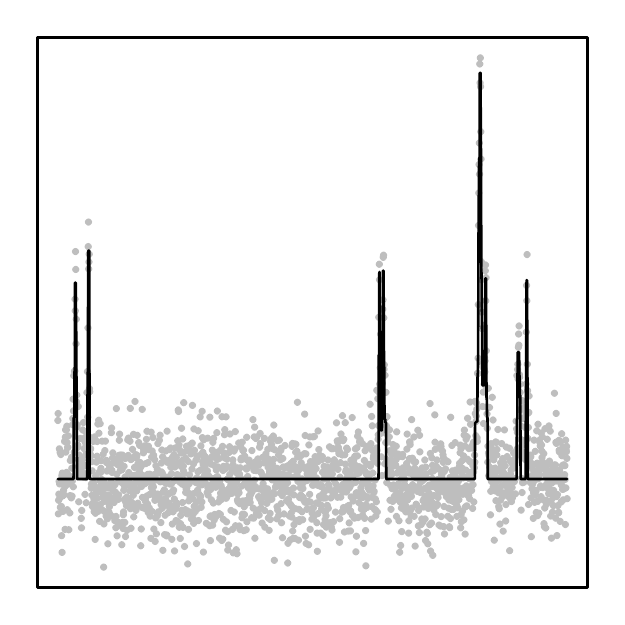}
\caption{\label{baseline}Scatterplots of data from spectroscopy. 
The solid lines show the function fitted by means of a total variation penalty (left) and the improved estimate of the baseline (right).}
\end{center}
\end{figure}

The data shown in Figure~\ref{baseline} are an excerpt from the spectroscopic analysis of a gallstone. Looking at the data, it seems reasonable to think of the points as having been generated by a function that is a flat baseline with occasional spikes.
Furthermore we have information about the correct location and number of spikes (Davies and Kovac 2001).

The left-hand plot in Figure~\ref{baseline} shows an estimate obtained by minimising (\ref{eq:tvdenoising}).
The smoothing parameters $\lambda_1,\ldots,\lambda_{n-1}$ were chosen by local squeezing, which aims to arrive at the smoothest function that satisfies the multiresolution condition
The smoothing parameters are only reduced in intervals where the multiresolution condition is not satisfied.
The estimates also show a mean correction: after running our algorithm we reset $f_i$ to the mean of the observations in $\mathcal{R}(i)$, for all $i$.
See Davies and Kovac (2001) for more details.

The estimate in Figure~\ref{baseline} identifies all the spikes.
However the left-hand estimate has not identified the constant baseline well.
Outside of the spikes, at the flat parts of the estimate, the fitted function takes many different values.

We propose a different graph that enables the algorithm to find a better estimate of the constant baseline.
We introduce a new vertex, indexed $n+1$.
This corresponds to a dummy observation with value $y_{n+1} = 0$.
We set the weight $w_{n+1} = 0$, so that the value of $y_{n+1}$ cannot influence the fitted function $f_1,\ldots,f_n$.
This new vertex is connected to the rest of the graph with $n$ new edges.
One new edge connects each existing observation to the dummy observation.

The idea is that the baseline regions (those observations or vertices that are not at a spike) will be joined together via the dummy vertex.
All of the baseline regions can be joined into one region.
The result is a constant baseline everywhere that there is not a bump.
The estimate of the baseline value will also improve, since the region contains more observations.

It is assumed that there are more observations in the baseline region than at a spike, so the dummy vertex will join the baseline region and not another region.

It remains to fix the values $\lambda_{1,n+1},\ldots,\lambda_{n,n+1}$. 
With no prior knowledge about the location of the spikes, we set $\lambda_{1,n+1} = \cdots = \lambda_{n,n+1} = \lambda_b$.
By using equal smoothing parameters we will not encourage any particular vertex to join the baseline region.
The other smoothing parameters, $\lambda_1,\ldots,\lambda_{n-1}$ are still chosen by local squeezing.
We suggest setting $\lambda_b = \min(\lambda_1,\ldots,\lambda_{n-1})$ so that no vertex will be influenced by the baseline more than its neighbours.

It is easy to see, in the right-hand plot of Figure~\ref{baseline}, the improvement that this graph causes at the baseline.

\subsection{Image analysis}
\begin{figure}
\begin{center}
\includegraphics{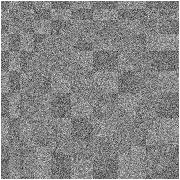}
\includegraphics{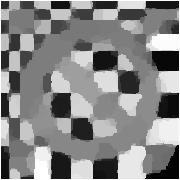}
\caption{\label{image}Noisy (left) and denoised (right) versions of the image of Polzehl and Spokoiny (2000).}
\end{center}
\end{figure}

Figure~\ref{image} shows, on the left, a noisy image that was used as an example by Polzehl and Spokoiny (2000).
This example demonstrates the use of our algorithm in the case where the graph is $(\mathcal{V}_4,\mathcal{E}_4)$, which is suggested by the 4-neighbourhood.

This particular image exhibits areas of solid colour, with sharp discontinuities between them.
We would expect to see this in many images.
Our algorithm works well on this kind of image, because the areas of solid colour can be represented by regions of constant value.

There are many proposed methods for choosing the smoothing parameters.
As, at this point, we are only interested in demonstrating our algorithm, we have employed a simple method suggested by Rudin et al.\ (1992).
It uses a global smoothing parameter, $\lambda$, and is based around an estimate of the global variance, $\sigma^2$.
Of course our algorithm allows different smoothing parameters at every edge, so we can make use of more complicated methods if we wish.

In order to find the simplest image for which the residuals behave as expected, we increase $\lambda$ until $\sum_{i \in \mathcal{V}_4} (f_i-y_i)^2 = \sigma^2 |\mathcal{V}_4|$. 
According to Chambolle (2004) this value of $\lambda$ will always exist.

Of course we require an estimate of $\sigma^2$ that is independent of the residuals. We can use, for example, one similar to that proposed by Davies and Kovac (2001):
\begin{displaymath}
\sigma = \frac{1.48}{\sqrt{2}} \mathop{\mathrm{median}} \left( |y_j - y_i| : (i,j) \in \mathcal{E}_4 \right).
\end{displaymath}

The output of our algorithm, the image estimated by use of the graph $(\mathcal{V}_4, \mathcal{E}_4)$, is shown in the right-hand image of Figure~\ref{image}.

\subsection{Irregularly-spaced data}
\begin{figure}[!t]
\begin{center}
\includegraphics[width=2.5in]{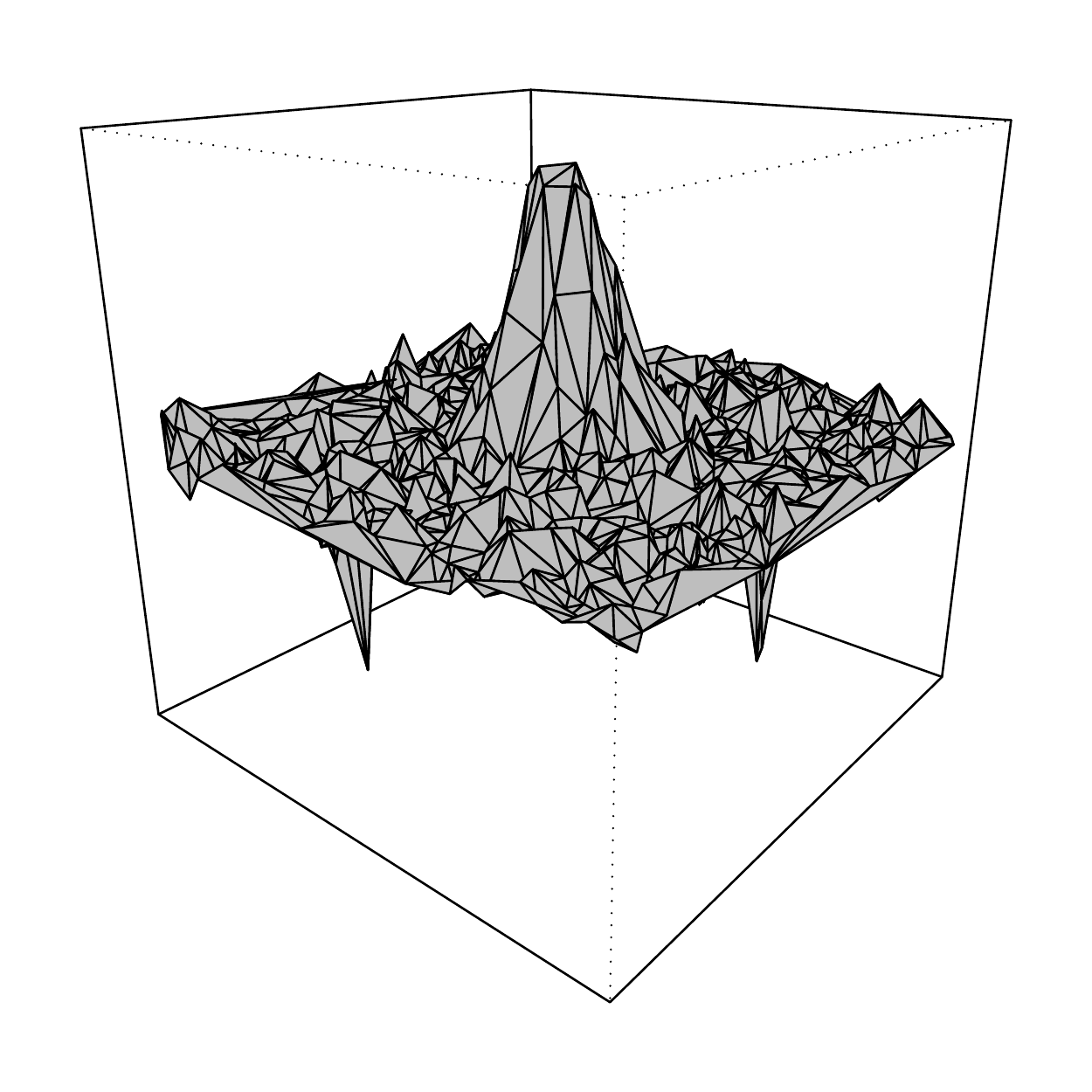}
\includegraphics{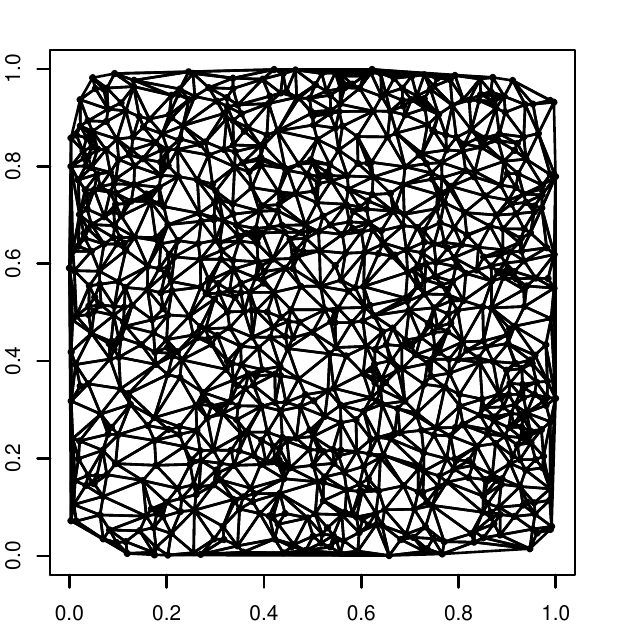} \\
\includegraphics[width=2.5in]{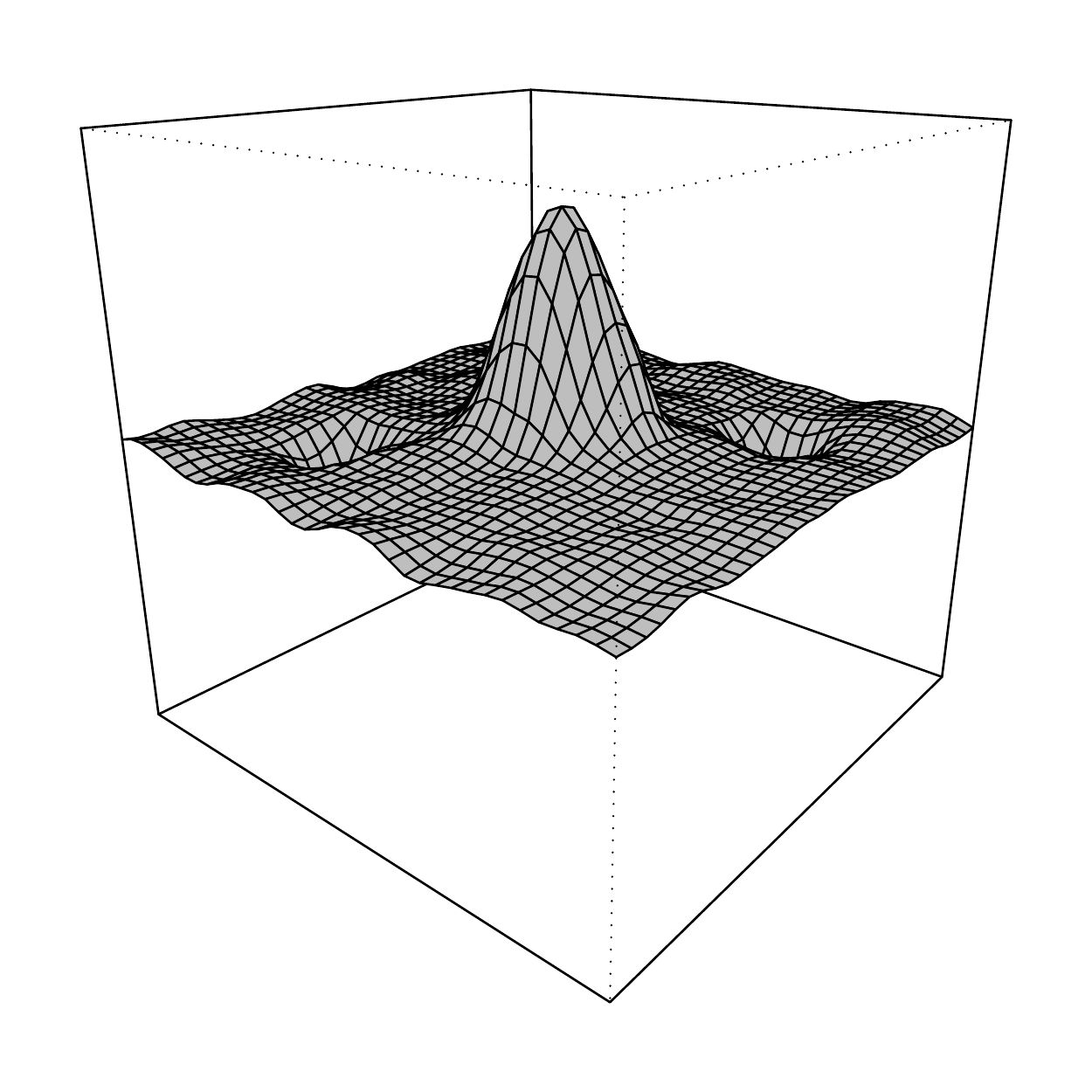}
\includegraphics[width=2.5in]{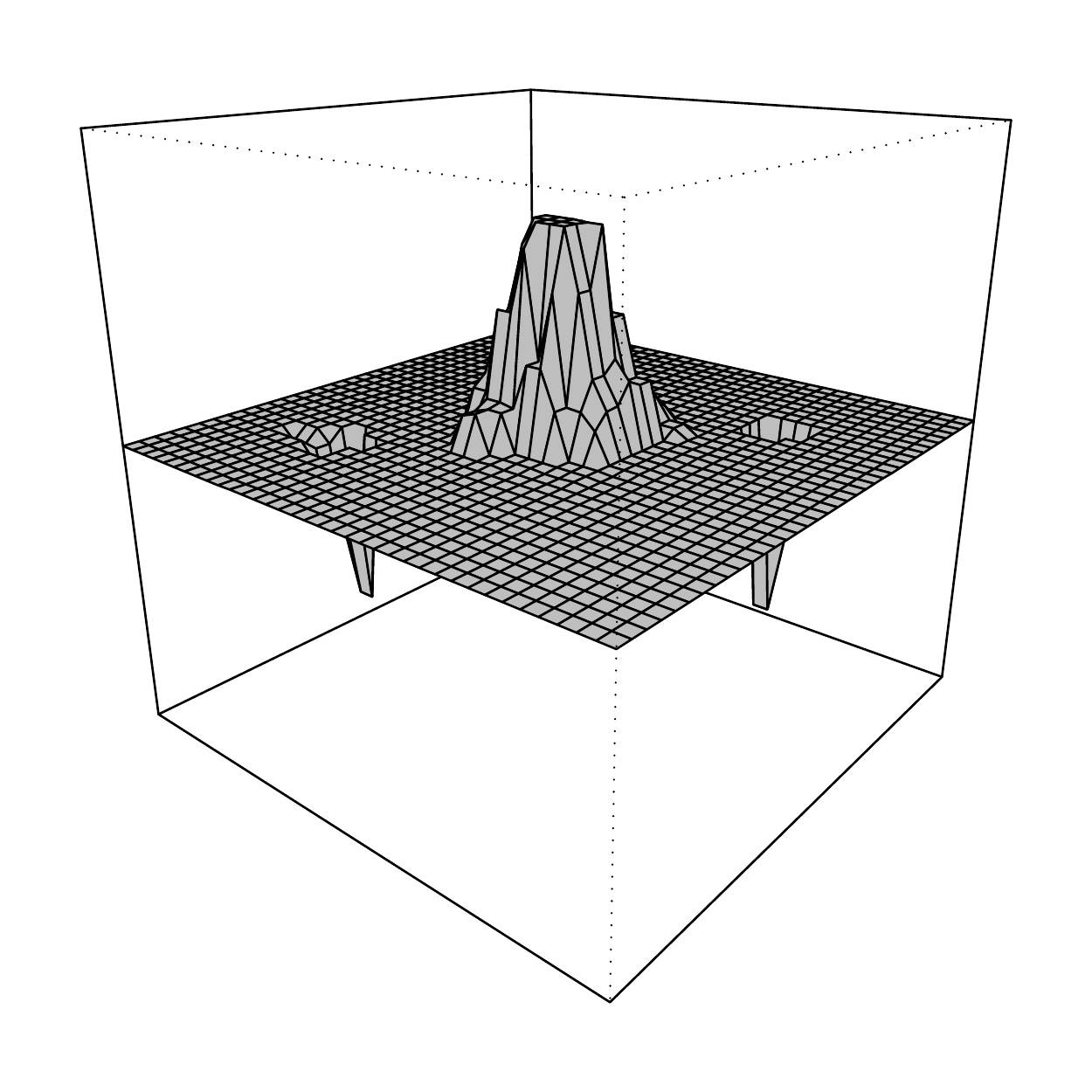}
\caption{\label{tri}Example of irregularly-spaced data.
The noisy simulated data is shown (top left) together with a kernel estimate (bottom left).
Note the presence of many additional bumps in the kernel estimate.
The Delaunay triangulation (top right) shows the location of vertices, and the edges of the graph that we obtain. The final plot (bottom right) shows the estimate obtained by minimising $Q(f)$ over the vertices of this graph.
}
\end{center}
\end{figure}

We generated 1000 covariates uniformly on $[0,1] \times [0,1]$. At each of these points we calculated a value from the function
\begin{eqnarray}
f(x_1, x_2) & = & \exp ( -100 ( (x_1 - 0.5)^2 + (x_2 - 0.5)^2) ) \nonumber \\
&  & {} - \exp ( -1000 ( (x_1 - 0.25)^2 + (x_2 - 0.25)^2) ) \nonumber \\
&  & {} - \exp ( -1000 ( (x_1 - 0.75)^2 + (x_2 - 0.75)^2) ). \label{eq:dfunc}
\end{eqnarray}
This function describes a surface with a broad bump at $(0.5,0.5)$ and two sharper, inverted bumps at $(0.25,0.25)$ and $(0.75,0.75)$.
To each of these values we added Gaussian noise with zero mean and standard deviation 0.05 to make 1000 noisy response observations.
The noisy surface is shown in Figure~\ref{tri}.

In order to calculate an estimate for $f$ the Delaunay triangulation was used to connect the irregularly spaced
covariates by a graph, see Figure~\ref{tri}.

For the sake of comparison, Figure~\ref{tri} also shows a kernel estimate applied to the data. We chose the global bandwidth that minimises the true squared error between the kernel estimate and the function given by (\ref{eq:dfunc}). So this can be thought of as the `best' global-bandwidth kernel estimate. Although it identifies the three bumps, it also exhibits many additional bumps in locations where the signal function is practically flat.

The bottom right plot in Figure~\ref{tri} shows the output of our algorithm, the result of minimising $Q(f)$ on the graph given by the Delaunay triangulation. We chose a global smoothing parameter by the same method as the image analysis example. This estimate identifies the three signal bumps but does not suffer from the introduction of extra bumps. There is a large region of constant value where the signal function is flat, so the estimate is also flat in these locations.

\appendix
\renewcommand{\theequation}{A.\arabic{equation}}
\setcounter{equation}{0}
\section{Appendix: Proofs}
\subsection{Proof of Theorem~\ref{theorem:one}}
We will show that (\ref{eq:entirecondition}), (\ref{eq:weakcondition}) and (\ref{eq:modifiedsplit}) are sufficient for $f$ to minimize $Q(f;c)$. Theorem~\ref{theorem:one} easily follows when (\ref{eq:stoppingrule}) also holds.

The problem of minimising $Q(f;c)$ can be posed as a constrained optimization problem with objective function
\begin{displaymath}
\frac{1}{2} \sum_{i \in \mathcal{V}} w_i (f_i - y_i)^2 + \sum_{(i,j) \in \mathcal{E}} c_{i,j} \lambda_{i,j} (f_j - f_i) - 2 \sum_{(i,j) \in \mathcal{E}} c_{i,j} \lambda_{i,j} v_{i,j},
\end{displaymath}
minimized subject to $c_{i,j} \lambda_{i,j} v_{i,j} \leq c_{i,j} \lambda_{i,j} (f_j - f_i)$ and $c_{i,j} \lambda_{i,j} v_{i,j} \leq 0$ for all $(i,j) \in \mathcal{E}$.

The Karush--Kuhn--Tucker conditions (see for example Bazaraa, Sherali and Shetty 1993, chap.\ 4) give a sufficient condition for $f$ and $v$ to be a solution. 
We require the existence of Lagrange multipliers $\mu_{i,j} \geq 0$ and $\mu_{i,j}' \geq 0$ such that $\mu_{i,j} = 0$ if $c_{i,j} v_{i,j} < c_{i,j} (f_j - f_i)$ and $\mu_{i,j}' = 0$ if $v_{i,j} \neq 0$, where $2 c_{i,j} \lambda_{i,j} = c_{i,j} \lambda_{i,j} \mu_{i,j} + c_{i,j} \lambda_{i,j} \mu_{i,j}'$ and
\begin{eqnarray}
\lefteqn{w_i(f_i - y_i) - \sum_{j : (i,j) \in \mathcal{E}} c_{i,j} \lambda_{i,j} + \sum_{j : (j,i) \in \mathcal{E}} c_{j,i} \lambda_{j,i}} \nonumber \\
& = & - \sum_{j : (i,j) \in \mathcal{E}} c_{i,j} \lambda_{i,j} \mu_{i,j} + \sum_{j : (j,i) \in \mathcal{E}} c_{j,i} \lambda_{j,i} \mu_{i,j}, \quad i \in \mathcal{V}. \label{eq:lagrange}
\end{eqnarray}
When (\ref{eq:weakcondition}) holds $c_{i,j} (f_j - f_i) > 0 \geq c_{i,j} v_{i,j}$ and hence $\mu_{i,j} = 0$ if $f_j \neq f_i$. Otherwise the non-negativity requirements on $\mu_{i,j}$ and $\mu_{i,j}'$ imply $0 \leq \mu_{i,j} \leq 2$.

Now suppose there exists an active set $\mathcal{A}$ such that $(\mathcal{V}, \mathcal{A})$ is acyclic, and (\ref{eq:entirecondition}) and (\ref{eq:modifiedsplit}) hold. The system of equations in (\ref{eq:lagrange}) is equivalent to the system of equations obtained by summing (\ref{eq:lagrange}) over all regions and subregions defined by $\mathcal{A}$. This system is: for every $a = k \in \mathcal{V}$ or $a = (I,J) \in \mathcal{A}$,
\begin{displaymath}
u_a f_l - m_a = - \sum_{i \in \mathcal{R}(a)} \left( \sum_{j : (i,j) \in \mathcal{E}} c_{i,j} \lambda_{i,j} \mu_{i,j} - \sum_{j : (j,i) \in \mathcal{E}} c_{j,i} \lambda_{j,i} \mu_{i,j} \right), \quad l \in \mathcal{R}(a),
\end{displaymath}
When (\ref{eq:entirecondition}) and (\ref{eq:modifiedsplit}) hold appropriate Lagrange multipliers exist for the above system of equations to be sufficient for $f$ to minimize $Q(f;c)$. Namely $\mu_{I,J} = -(u_{I,J} f_I - m_{I,J}) / c_{I,J} \lambda_{I,J}$ if $(I,J) \in \mathcal{A}$ and $\mu_{I,J} = 0$ otherwise.

\subsection{Alterations of the active set}
In this subsection we prove the different values of $\Delta f_k$, $\Delta f_l$ and $\Delta c_{k,l}$ associated with the events described in Subsection \ref{section:algorithm}.

The condition (\ref{eq:entirecondition}) tells us that $u_k f_k = m_k$ and $u_l f_l = m_l$, and also $u_k(f_k + \Delta f_k) = m_k + \Delta c_{k,l} \lambda_{k,l}$ and $u_l(f_l + \Delta f_l) = m_l - \Delta c_{k,l} \lambda_{k,l}$.
Combining these equations we see that we must have
\begin{eqnarray}
u_k \Delta f_k & = & \phantom{-} \Delta c_{k,l} \lambda_{k,l}, \label{eq:1k} \\
u_l \Delta f_l & = & -\Delta c_{k,l} \lambda_{k,l}, \label{eq:1l} 
\end{eqnarray}

\subsubsection{No change to active set}
If there are no necessary changes to the active set, then $c_{k,l}$ will reach the target value of $\mathrm{\mathop{sign}}(f_l - f_k)$. 
Therefore $\Delta c_{k,l} = \mathrm{\mathop{sign}}(f_l - f_k) - c_{k,l} \neq 0$.
The values of $\Delta f_k$ and $\Delta f_l$ follow from (\ref{eq:1k}) and (\ref{eq:1l}) respectively, as does the requirement that $u_k > 0$ and $u_l > 0$.

\subsubsection{Merging of the two regions}
The two regions $\mathcal{R}(k)$ and $\mathcal{R}(l)$ will merge when $f_k + \Delta f_k = f_l + \Delta f_l$. 
Provided that $u_k > 0$ and $u_l > 0$, combining the above equation with (\ref{eq:1k}) and (\ref{eq:1l}) gives (\ref{eq:merge}).
If $u_k = u_l = 0$ then we can set $f_k = f_l$ equal to any value that we choose, such as the mean and median value $(f_k + f_l) / 2$.

\subsubsection{Amalgamation of a neighbouring region}
Given a suitable vertex $K$, the two regions $\mathcal{R}(k)$ and $\mathcal{R}(K)$ will amalgamate when $f_k + \Delta f_k = f_K$. 
When $u_l > 0$ the values in (\ref{eq:amalg}) follow immediately from (\ref{eq:1k}) and (\ref{eq:1l}).
When $u_l = 0$ equating (\ref{eq:1k}) and (\ref{eq:1l}) shows that either $u_k = 0$ or $\Delta f_k = 0$. 
In either case it makes little sense to alter $f_l$, so we let $\Delta f_l = 0$. 

\subsubsection{Splitting a region}
Suppose we split $\mathcal{R}(k)$ by removing $(I,J)$ from $\mathcal{A}$.
The value of $f_k$ at which this happens satisfies (\ref{eq:modifiedsplit}) in equality.
Without loss of generality suppose $k \in \mathcal{R}(I,J)$. We will need to swap the sign of $c_{I,J}$ if $\mathop{\mathrm{sign}}(c_{I,J}) = \mathop{\mathrm{sign}}(\Delta f_k) = \mathop{\mathrm{sign}}(f_l - f_k)$. Once this is taken into account $m_{I,J}$ becomes $m_{I,J} - c_{I,J} \lambda_{I,J} - \mathop{\mathrm{sign}}(f_l - f_k) |c_{I,J}| \lambda_{I,J}$ and (\ref{eq:modifiedsplit}) becomes
\begin{displaymath}
0 \leq |\Delta f_k| + \mathop{\mathrm{sign}}(f_l - f_k) (u_{I,J} f_k - m_{I,J} + c_{I,J} \lambda_{I,J}) + |c_{I,J}| \lambda_{I,J} \leq 2 |c_{I,J}| \lambda_{I,J}.
\end{displaymath}
The value for $\Delta f_k$ follows when the upper limit is satisfied in equality.
If $u_k = 0$ then $u_{I,J}$ = 0 so (\ref{eq:modifiedsplit}) will never change when $f_k$ changes.
If $u_l = 0$ then $\Delta f_k = 0$ from equating (\ref{eq:1k}) and (\ref{eq:1l}).
Clearly for $f_k$ to change and a split to occur we must have $u_k > 0$ and $u_l > 0$.
The value for $\Delta f_l$ follows from equating (\ref{eq:1k}) and (\ref{eq:1l}).

\subsection{Proof of Theorem~\ref{theorem:two}}
We will show that the objective function at the end of each iteration, $Q(f + \Delta f ; c + \Delta c)$, is never less than the objective function at the start of the iteration, $Q(f;c)$. Since $f$ minimises $Q(f;c)$ and $\Delta c_{i,j} = 0$ except for $(i,j) = (k,l)$, we have
\begin{eqnarray*}
& & Q(f + \Delta f ; c + \Delta c)  - Q(f;c) \\
& = & Q(f + \Delta f ; c) - Q(f;c) + |\Delta c_{k,l}| \lambda_{k,l} |f_l + \Delta f_l - f_k - \Delta f_k| \\
& \geq & |\Delta c_{k,l}| \lambda_{k,l} |f_l + \Delta f_l - f_k - \Delta f_k| \geq 0.
\end{eqnarray*}
Equality can only occur when $\Delta c_{k,l} = 0$ or $f_l + \Delta f_l = f_k - \Delta f_k$. So the only time that $Q(f;c)$ does not increase is during merging or amalgamation.
Therefore an edge cannot be removed from the active set without an increase in $Q(f;c)$. This means that the algorithm never visits the same value of $c$ and $\mathcal{A}$ twice, and will always arrive at the situation described in (\ref{eq:stoppingrule}) and terminate.

\section*{References}
\begin{description}
\item Bazaraa, M. S., Sherali, H. D., and Shetty, C. M. (1993), \emph{Nonlinear Programming}, New York: John Wiley \& Sons.
\item Belkin, M., Matveeva, I., and Niyogi, P.\ (2004), ``Regularization and Semi-supervised Learning on Large Graphs,'' in \emph{Learning Theory}, eds. J.\ Shawe-Taylor and Y.\ Singer, Berlin: Springer-Verlag, pp.\ 624--638.
\item Chambolle, A. (2004), ``An Algorithm for Total Variation Minimization and Applications,'' \emph{Journal of Mathematical Imaging and Vision}, \textbf{20}, 89--97.
\item Davies, P. L., and Kovac, A. (2001), ``Local Extremes, Runs, Strings and Multiresolution,'' \emph{The Annals of Statistics}, \textbf{29}, 1--65.
\item Goldfarb, D., and Idnani, A. (1983), ``A Numerically Stable Dual Method for Solving Strictly Convex Quadratic Programs,'' \emph{Mathematical programming}, \textbf{27}, 1--33.
\item Jansen, M., Nason, G. P., and Silverman, B. W. (2009), ``Multiscale Methods for Data on Graphs and Irregular Multidimensional Situations,'' \emph{Journal of the Royal Statistical Society, Series B}, \textbf{71}, 97--125.
\item Koenker, R., and Mizera, I. (2004), ``Penalized Triograms; Total Variation Regularization for Bivariate Smoothing,'' \emph{Journal of the Royal Statistical Society, Series B}, \textbf{66}, 145--163.
\item Mammen, E., and van de Geer, S. (1997), ``Locally Adaptive Regression Splines,'' \emph{The Annals of Statistics}, \textbf{25}, 387--413.
\item Polzehl, J., and Spokoiny, V. G. (2000), ``Adaptive Weights Smoothing With Applications to Image Restoration,'' \emph{Journal of the Royal Statistical Society, Series B}, \textbf{62}, 335--354.
\item Rudin, L. I., Osher, S., and Fatemi, E. (1992), ``Nonlinear Total Variation Based Noise Removal Algorithms,'' \emph{Physica D}, \textbf{60}, 259--268.
\item Tibshirani, R. (1996), ``Regression Shrinkage and Selection via the Lasso,'' \emph{Journal of the Royal Statistical Society, Series B}, \textbf{58}, 267--288.
\item Winkler, G. (2003), \emph{Image Analysis, Random Fields and Markov Chain Monte Carlo Methods}, Berlin: Springer-Verlag.
\end{description}

\end{document}